\documentclass[reprint, superscriptaddress, secnumarabic, amssymb, nobibnotes, aps, prl]{revtex4-1}

\usepackage{graphicx}
\usepackage{epstopdf}
\usepackage[T1]{fontenc}
\usepackage[latin9]{inputenc}
\usepackage{amsbsy}
\usepackage{gensymb}
\usepackage[T1]{fontenc}
\usepackage[latin9]{inputenc}
\usepackage{amsmath}
\usepackage{amssymb}
\usepackage{bbm}
\usepackage{braket}
\usepackage{xcolor}
\allowdisplaybreaks
\usepackage{graphicx}
\usepackage[colorlinks=true]{hyperref}  
\hypersetup{
    bookmarks=true,         
    unicode=false,          
    pdftoolbar=true,        
    pdfmenubar=true,        
    pdffitwindow=false,     
    pdfstartview={FitH},    
    pdftitle={x},    
    pdfauthor={SM},     
    pdfsubject={},   
    pdfcreator={},   
    pdfproducer={}, 
    pdfkeywords={} {} {}, 
    pdfnewwindow=true,      
    colorlinks=true,       
    linkcolor=blue, 
    citecolor=blue,        
    filecolor=magenta,      
    urlcolor=blue           
} 
\usepackage[normalem]{ulem}

\newcommand{\figref}[1]{Fig.~\ref{#1}}

\renewcommand{\approx}{\simeq}

\begin{document}
\title{\textrm{Superconductivity in a new hexagonal high entropy alloy}}
\author{Sourav Marik}
\affiliation{Indian Institute of Science Education and Research Bhopal, Bhopal, 462066, India}
\affiliation{Laboratoire CRISMAT, UMR 6508 CNRS ENSICAEN, 6 bd du Marechal Juin, 14050 Caen Cedex 4, France}
\author{Kapil Motla}
\affiliation{Indian Institute of Science Education and Research Bhopal, Bhopal, 462066, India}
\author{Maneesha Varghese}
\affiliation{Indian Institute of Science Education and Research Bhopal, Bhopal, 462066, India}
\author{K. P. Sajilesh}
\affiliation{Indian Institute of Science Education and Research Bhopal, Bhopal, 462066, India}
\author{Deepak Singh}
\affiliation{Indian Institute of Science Education and Research Bhopal, Bhopal, 462066, India}
\author{Y. Breard}
\affiliation{Laboratoire CRISMAT, UMR 6508 CNRS ENSICAEN, 6 bd du Marechal Juin, 14050 Caen Cedex 4, France}
\author{P. Boullay}
\affiliation{Laboratoire CRISMAT, UMR 6508 CNRS ENSICAEN, 6 bd du Marechal Juin, 14050 Caen Cedex 4, France}
\author{R. P. Singh}
\email[]{rpsingh@iiserb.ac.in}
\affiliation{Indian Institute of Science Education and Research Bhopal, Bhopal, 462066, India}

\date{\today}
\begin{abstract}
\begin{flushleft}

\end{flushleft}
High entropy alloys (HEAs) are the new class of materials with an attractive combination of tunable mechanical and physicochemical properties.  They crystallize mainly in cubic structures, however, for practical applications, HEAs with hexagonal close-packed (hcp) structure are highly desirable in connection to their in general high hardness. Herein, we report the synthesis, structure and detailed superconducting properties of Re$_{0.56}$Nb$_{0.11}$Ti$_{0.11}$Zr$_{0.11}$Hf$_{0.11}$-the first hexagonal superconducting high entropy alloy (HEA) composed of five randomly distributed transition-metals. Combination of room temperature precession electron diffraction, precession electron diffraction tomography  and powder X-ray diffraction is utilized to determine the room temperature crystal structure.  Transport, magnetic and heat capacity measurements show that the material is a type-II superconductor with the bulk superconducting transition at $T_{c}$ = 4.4 K, lower critical field $H_{c1}$(0) = 2.3 mT and upper critical field $H_{c2}$(0) = 3.6 T. Low-temperature specific heat measurement indicates that Re$_{0.56}$Nb$_{0.11}$Ti$_{0.11}$Zr$_{0.11}$Hf$_{0.11}$ is a phonon-mediated superconductor in the weak electron-phonon coupling limit with a normalized specific heat jump $\frac{\Delta C_{el}}{\gamma_{n}T_{c}}$ = 1.32. Further, hexagonal to cubic structural transition is observed by lowering the valence electron counts and $T_{c}$ follows crystalline-like behaviour. 
\end{abstract}
\maketitle
\section{Introduction}

Alloys with superior functional and mechanical properties are among the most germane materials for modern technologies. In this context, the advent of a new class of materials termed as high entropy alloys (HEAs) have drawn intense interest in recent times and has initiated a fascinating interdisciplinary research topic \cite{1,2,3,4,5}. HEAs are constructed from five or more principal elements in equimolar or near equimolar ratio. The high mixing entropic contribution ($\Delta$S$_{mixing}$) to the decreased Gibbs free energy stabilizes disordered solid solution phases in simple crystallographic lattices (such as body-centered (bcc) and face-centered cubic (fcc) structures) \cite{2,3,4,5}. Combining simple structures with a high chemical disorder, this novel design strategy enables to access the materials with properties superior to those of conventional alloys and phases. Illustrative examples include a superior mechanical performance at high and cryogenic temperatures \cite{6}, simultaneous strength and ductility \cite{7,8}, superior hydrogen storage capacity \cite{9}, high mechanical strength, corrosion resistance and thermal stability \cite{2,3,5}. At the same time, very recently, exotic quantum phenomena such as complex magnetism \cite{10} and superconductivity \cite{11,12,13,14,15,16,17} are discovered in HEAs. Among several studied HEAs, only a handful of HEAs show superconductivity and all of them possess simple cubic structure. Nb-Ti (Nb-Ti based alloys are still dominating as superconducting (SC) magnets, which are used in NMR, MRI devices) derived HEA [TaNb]$_{0.67}$[HfZrTi]$_{0.33}$ is the first reported HEA superconductor (bcc cubic structure) with a superconducting transition temperature ($T_{c}$) 7.3 K, and shows an upper critical field $H_{c2}$(0) of 8.2 T \cite{12,13,14}. [ScZrNb]$_{0.65}$[RhPd]$_{0.35}$ (cubic primitive CsCl-type lattice) shows superconductivity with $T_{c}$=9.3 K, and exhibits $H_{c2}$(0) = 10.7 T \cite{15}. Recently, superconductivity is reported in $\alpha$ - Mn type Re based HEAs \cite{16}. Further, the niobium-titanium derived HEA shows unique and extraordinary robust zero-resistance superconductivity at pressure up to 190 GPa and therefore, making HEA superconductor a promising candidate for the future superconducting applications \cite{18}.\

To date, a majority of the HEAs are reported to possess one of two simple crystal structures: face-centered cubic (fcc) or body-centered cubic. Nevertheless, such cubic HEAs prepared, to date, suffer from poor ductility at room temperature, and this     limits their practical applications. On the other hand, HEAs with the hex-agonal closed packed (hcp) structure, which is rare among HEAs is desirable for practical use in connection with their in general high hardness. Most of the reported hcp HEAs are composed of rare earth elements and contrary to the transition metal HEAs, these materials, due to the size and electronic structure mismatches among component elements lack the solid solution strengthening effect \cite{10,19,20,21}. There is a single known transition metal only HEA system, CrMnFeCoNi (non superconducting), which transform into a hcp phase under a high external pressure of 14 GPa and the hcp phase is retained following decompression to ambient pressure \cite{22}.\

In this paper, we not only report new HEA superconductors with cubic structure but also report the first hexagonal HEA superconductor, Re$_{0.56}$Nb$_{0.11}$Ti$_{0.11}$Zr$_{0.11}$Hf$_{0.11}$, exhibiting spectacular type-II superconductivity at $T_{c}$=4.4 K. The material is a mixed 4d 5d system and a rare example of transition metal only hexagonal high entropy alloy. Detailed structural and superconducting properties are described as determined by powder X-ray diffraction, electron microscopy, resistivity, magnetization, and specific heat measurements.

\section{Experimental Details}
\textbf{Synthesis.} Polycrystalline samples of nominal composition [Nb$_{0.67-x}$Re$_{x}$][TiZrHf]$_{0.33}$ were synthesized using the standard arc-melting technique. Stoichiometric amounts of Nb (purity 99.8 $\%$), Re (purity 99.99 $\%$), Zr (purity 99.95 $\%$), Hf (purity 99.7 $\%$) and Ti (purity 99.99 $\%$) pieces were taken on a water cooled copper hearth under the flow of high purity argon gas. They were melted in high current (T>2500) to make a single button, then flipped and re-melted several times for the sample homogeneity. The samples formed were hard with insignificant weight loss.

\textbf{X-ray Powder Diffraction and Scanning Electron Microscopy.} Powder X-ray diffraction (XRD) was carried out at room temperature (RT) on a PANalytical diffractometer equipped with Cu-K$_{\alpha}$ radiation ($\lambda$ = 1.54056 \text{\AA}). Rietveld refinements were performed using the JANA2006 program \cite{23}. Sample compositions were checked by a scanning electron microscope (SEM) equipped with an energy-dispersive spectrometer.

\textbf{Precession electron diffraction (PED) and precession electron diffraction tomography (PEDT).} RT PED and PEDT were performed with a JEOL 2010 electron microscope (200 kV, LaB6 cathode) equipped with a nanomegas DigiStar precession module and a Gatan Orius 200D CCD camera. PED patterns and PEDT datasets were recorded at room temperature on several crystals. The precession angle was set to 1.5 with a goniometer tilt step below 1 degree. Crystal symmetry and the diffraction data were analyzed using the PETS \cite{24} and JANA2006 \cite{23} softwares. Sample compositions were checked by energy dispersive X-ray spectroscopy (EDS) before recording the PEDT data. 

\textbf{Physical Property Measurements.} DC magnetization and ac susceptibility measurements were performed using a Quantum Design superconducting quantum interference device (MPMS 3, Quantum Design). The electrical resistivity measurements were performed on the physical property measurement system (PPMS, Quantum Design, Inc.) by using a conventional four-probe ac technique at frequency 17 Hz and excitation current 10 mA. The measurements were carried out under the presence of different magnetic fields. Specific heat measurement was performed by the two tau time-relaxation method using the PPMS in zero magnetic field.

\section{Results and Discussion}
\begin{figure}
\includegraphics[width=1.0\columnwidth]{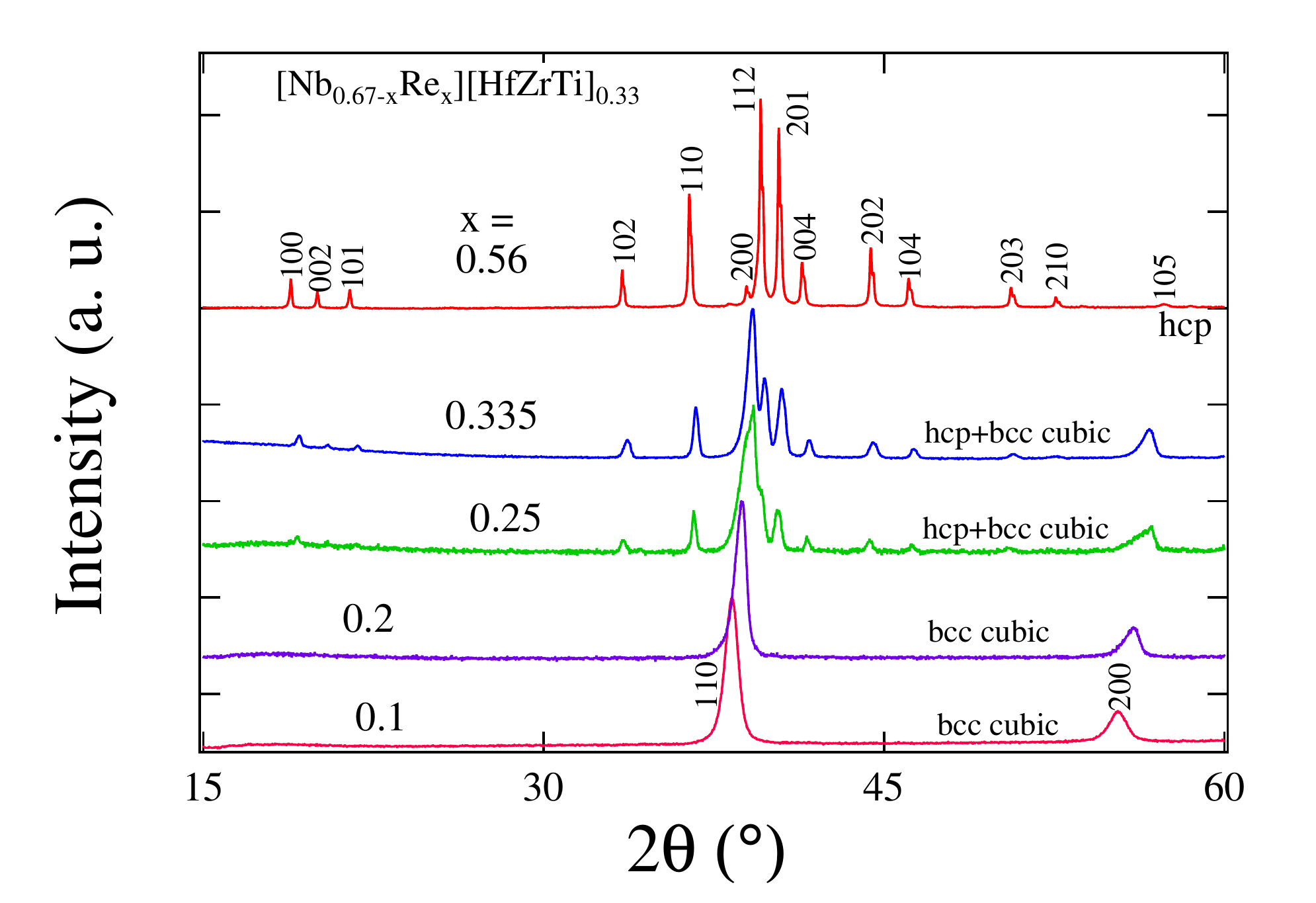}
\caption{\label{Fig1:XRD}Room temperature (RT) powder XRD patterns for the [Nb$_{0.67-x}$Re$_x$][TiZrHf]$_{0.33}$ materials. The sample with higher Re content (x = 0.56, e/a = 5.79) can be identified as pure phase and the XRD pattern can be indexed with a hcp structure (P6$_3$/mmc, a = b = 5.25 (1) and c = 8.59 (1) \text{\AA}). On the other hand, samples with lower Re content (lower e/a ratios) transform into a bcc cubic structure (Space Group Im -3m).}
\end{figure}

\begin{figure}
\includegraphics[width=1.0\columnwidth]{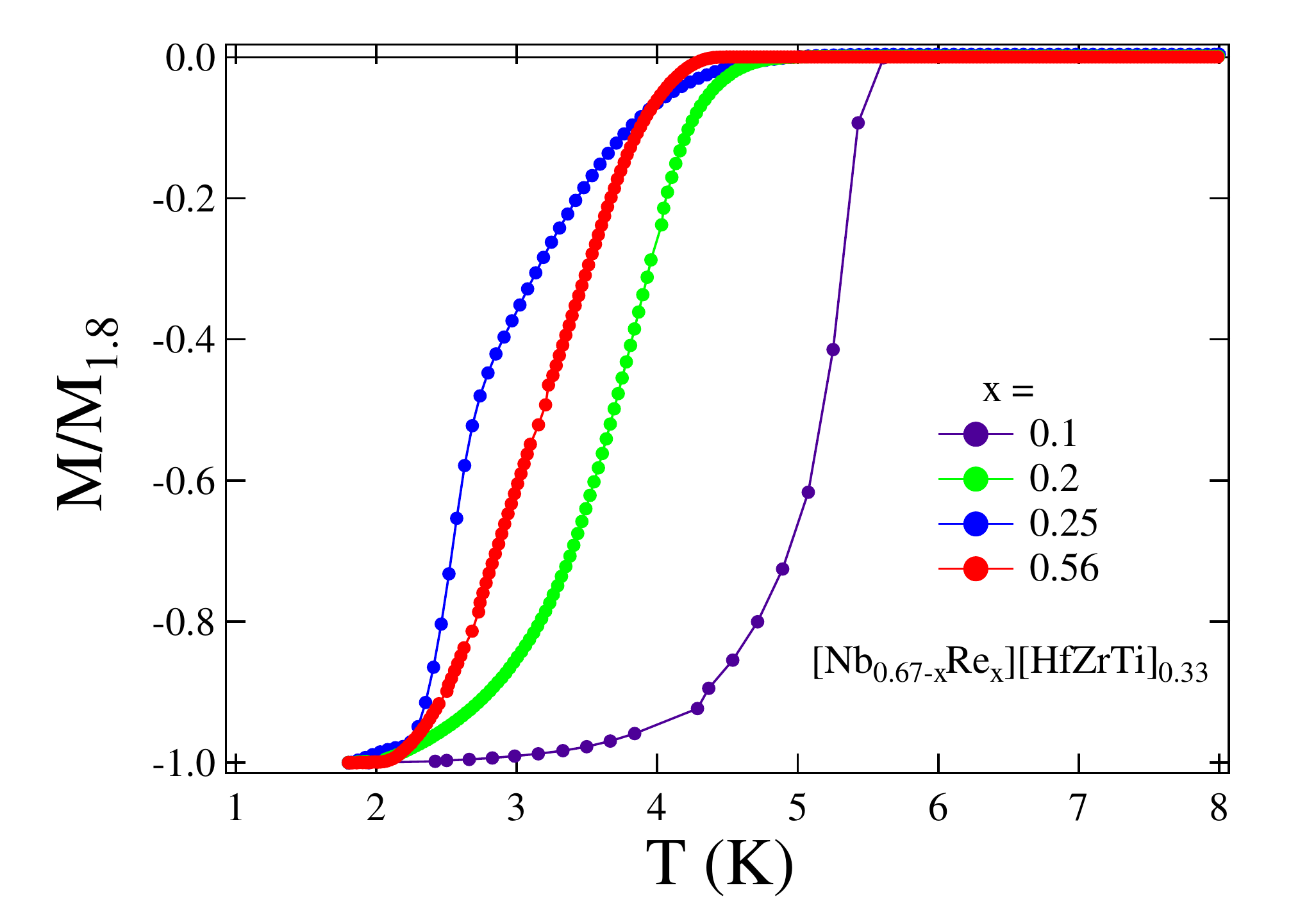}
\caption{\label{Fig2:ALLMT}Temperature variation of the normalized magnetic moment for the [Re$_{x}$Nb$_{0.67-x}$][TiZrHf]$_{0.33}$ materials taken in 1 mT field. Double transition is observed for the sample with x = 0.25, this in fact supports the observation of mixed phase (hcp+bcc) in the RT XRD pattern of the same material.}
\end{figure}

\textbf{Structural characterization and superconducting behaviour of [Nb$_{0.67-x}$Re$_{x}$][TiZrHf]$_{0.33}$ compounds.} \figref{Fig1:XRD} shows the Room temperature (RT) powder X-ray diffraction (XRD) pattern for the [Nb$_{0.67-x}$Re$_{x}$][TiZrHf]$_{0.33}$ materials. Samples with lower Re content shows cubic symmetry (Space Group Im -3m). However, a structural transition from cubic (bcc) to hexagonal is observed with increasing Re content in the structure. The sample with higher Re content (x = 0.56, e/a = 5.79) can be identified as pure phase and the XRD pattern can be indexed with a hcp structure (S.G. P6$_3$/mmc, a = b = 5.25 (1) and c = 8.59 (1) \text{\AA}). In the case of conventional alloys, a structural transition from bcc to hcp is generally observed with increasing electron count (e/a). We have observed this polymorphic transition with increasing Re (increasing e/a ratio) content in the structure. This structural transition begins at e/a = 5.17, then with lowering the e/a ratio the structure transforms into a cubic one. It is noteworthy that a similar structural transformation (cubic to hcp) is observed for the transition metal only CrMnFeCoNi system, however under a high external pressure of 14 GPa \cite{22}. Fig. \ref{Fig2:ALLMT} shows the temperature variation of the dc susceptibility for all the [Nb$_{0.67-x}$Re$_x$][TiZrHf]$_{0.33}$ materials. An increase in the value of $T_{c}$ is observed with decreasing e/a ratio for the pure phase materials.\\ 

\begin{figure}
\includegraphics[width=1.0\columnwidth]{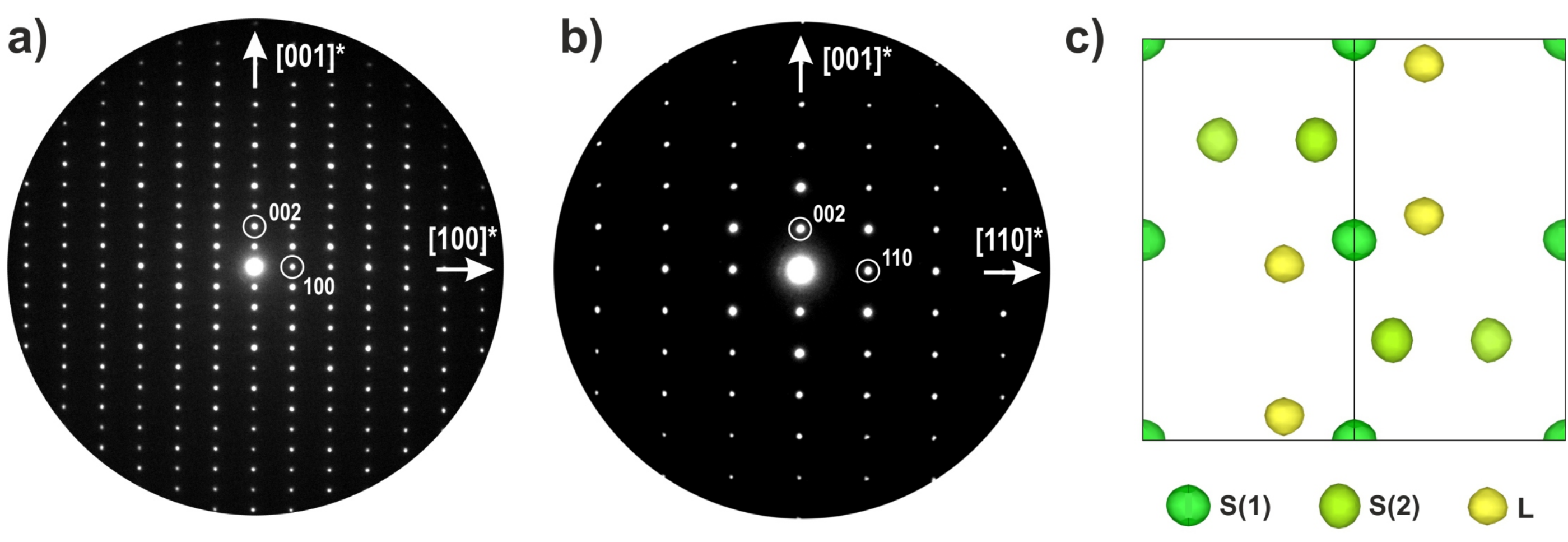}
\caption{\label{Fig3:PEDT}(a) [010] and (b) [1$\bar{1}$0] PED zone axis patterns obtained for Re$_{0.56}$Nb$_{0.11}$Ti$_{0.11}$Zr$_{0.11}$Hf$_{0.11}$. In (b), the condition {\it hhl: l=2n} is observed. (c) [110] projection of the density map resulting from the structure solution procedure indicating the existence of 3 atomic positions S(1), S(2) and L (see Table 1).}
\end{figure}

\begin{figure}
\includegraphics[width=1.0\columnwidth]{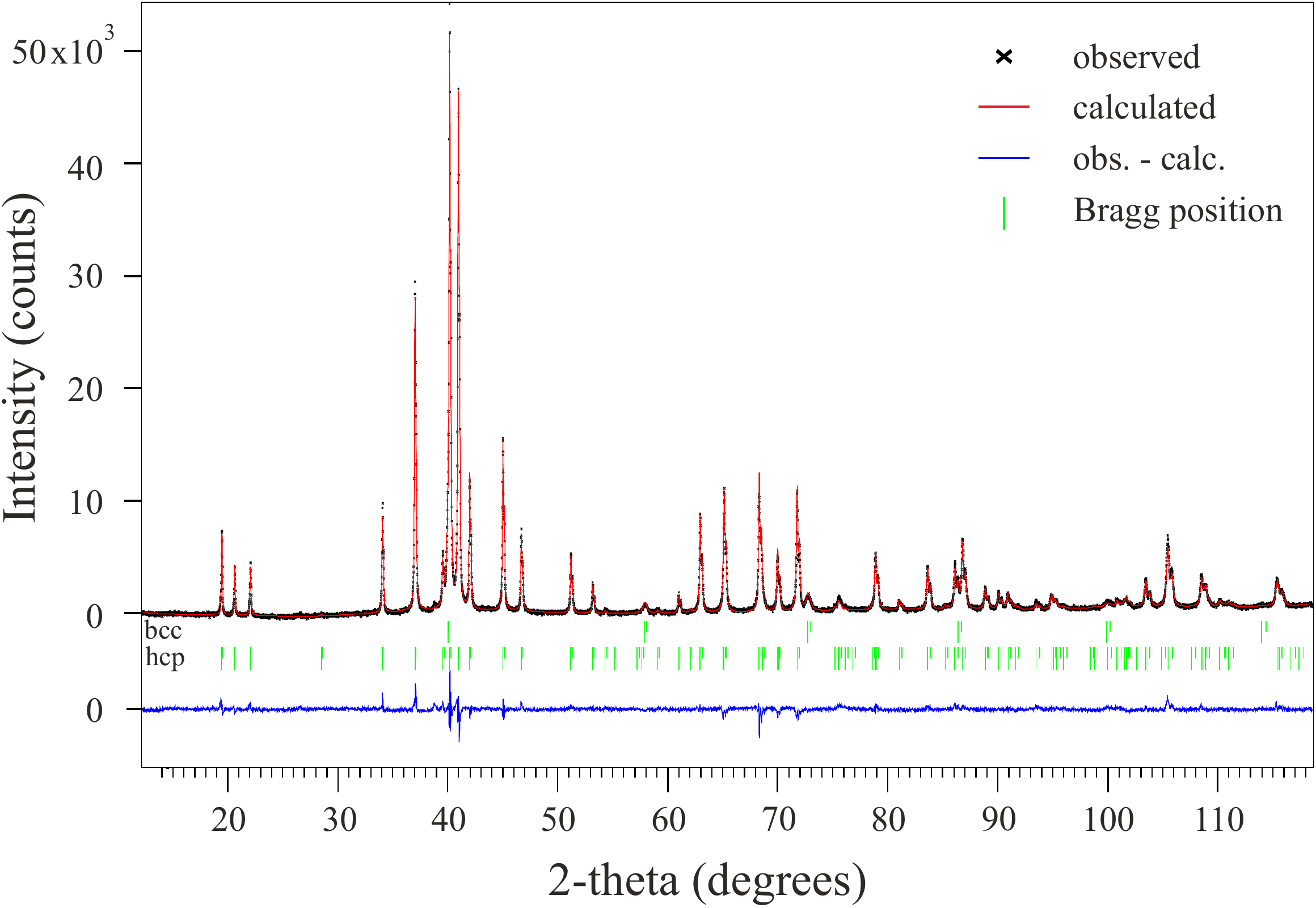}
\caption{\label{Fig4:RV}Final observed, calculated and difference plots obtained for the Rietveld refinement of the room temperature X-ray powder diffraction pattern of Re$_{0.56}$Nb$_{0.11}$Ti$_{0.11}$Zr$_{0.11}$Hf$_{0.11}$.}
\end{figure}
\begin{figure}
\includegraphics[width=1.0\columnwidth]{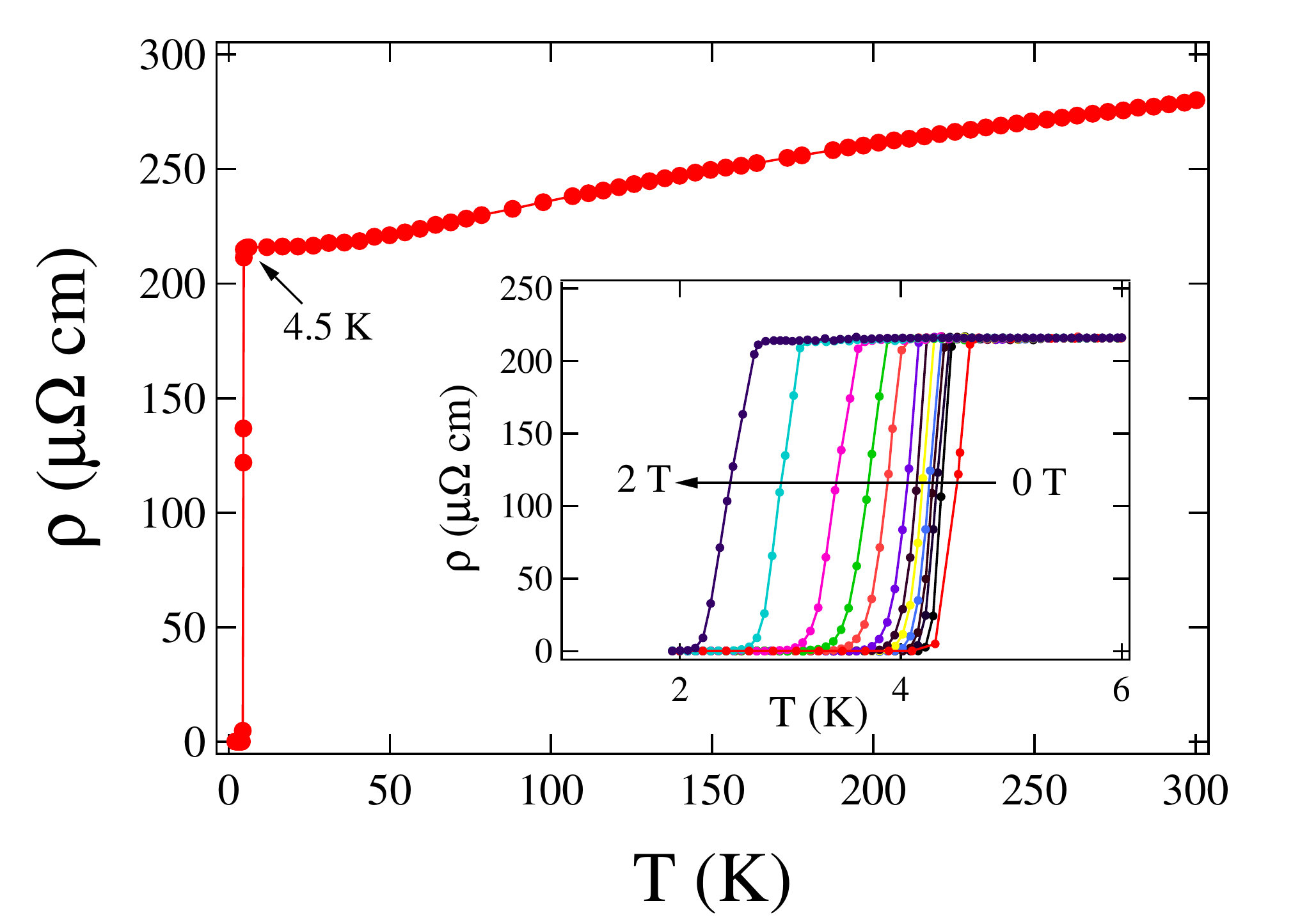}
\caption{\label{Fig5:RESISNO}Temperature dependence of resistivity under zero magnetic field for Re$_{0.56}$Nb$_{0.11}$Ti$_{0.11}$Zr$_{0.11}$Hf$_{0.11}$. Inset shows the magnetic field variation of the resistivity.}
\end{figure}

\textbf{Crystal structure determination and superconducting behaviour of hcp Re$_{0.56}$Nb$_{0.11}$Ti$_{0.11}$Zr$_{0.11}$Hf$_{0.11}$}. The PED patterns recorded on the Re$_{0.56}$Nb$_{0.11}$Ti$_{0.11}$Zr$_{0.11}$Hf$_{0.11}$ is shown in \figref{Fig3:PEDT}. These are compatible with the reflection conditions ({\it hhl: l=2n}, {\it 00l: l=2n}) observed for the P6$_3$/mmc space group (\figref{Fig3:PEDT} (a and b)). The structure solution based on PEDT data (not shown, using JANA2006) unambiguously confirms the hcp structure of the compound. The potential map and expected atomic positions resulting from this structure solution is highlighted in \figref{Fig3:PEDT} (c). The crystal structure is refined further using the Rietveld method using powder XRD pattern recorded at RT (Fig. \ref{Fig4:RV}). In {Re$_{0.56}$Nb$_{0.11}$Ti$_{0.11}$Zr$_{0.11}$Hf$_{0.11}$}, the element with the smallest covalent radii (Re) is preferentially located in the 2a (0,0,0) and 6h (x,2x,1/4) Wickoff positions (S(1) and S(2) in \figref{Fig3:PEDT} (c) and Table 1), while elements with the larger covalent radii (Hf and Zr) are located in the 4f (1/3,2/3,z) position, (L in \figref{Fig3:PEDT} (c) and Table 1). Nevertheless, the intermediate radii elements, Nb and Ti could occupy any of the three atomic positions. The possibility of Nb  shares atomic positions with Re (S(1) and S(2)) tends to result in a better refinement leading to the structural parameters shown in Table 1. 

\begin {table}
\caption {Refined structural parameters for hcp Re$_{0.56}$Nb$_{0.11}$Ti$_{0.11}$Zr$_{0.11}$Hf$_{0.11}$, Re and Nb occupy the S(1) and S(2) sites, and Ti, Zr and Hf occupy L sites.}  
\begin{center}
\begin{tabular}{ |c c c c c c| } 
 \hline
  & Wick. Pos. & x & y & z & Uiso (\text{\AA}$^{2}$) \\ 
 S(1) & 2a & 0 & 0 & 0 & 0.0038(2) \\ 
 S(2) & 6h & 0.17233(4) & 0.34466(8) & 1/4 & 0.01154(17) \\
 L & 4f & 2/3 & 1/3 & 0.0638(7) & 0.0083(2) \\ 
\hline
\multicolumn{6}{c}{a=5.2554(1)\text{\AA}, c=8.5934(1)\text{\AA}, R$_{P}$=1.48, R$_{WP}$=2.04, GOF=1.75} 
\end{tabular}
\end{center}
\end{table}
Fig. \ref{Fig5:RESISNO} shows the temperature variation of zero magnetic field resistivity, which shows a sharp drop to a zero-resistivity superconducting state below 4.5 K. The normal state resistivity shows a sluggish increment with temperature, highlighting a poor metallic behavior. The residual resistivity ratio (RRR) is found to be $\rho$(300)/$\rho$(10) = 1.2, which is very small, suggesting the existence of atomic- scale disorder in the HEA. However, the small RRR value can be compared to those of previously reported HEAs \cite{12,15,16}. The inset in Fig. \ref{Fig5:RESISNO} shows the temperature dependence of resistivity at different magnetic fields ($\rho$(H)-T). This illustrates a gradual suppression of $T_{c}$ with increasing magnetic field.
\begin{figure}
\includegraphics[width=1.0\columnwidth]{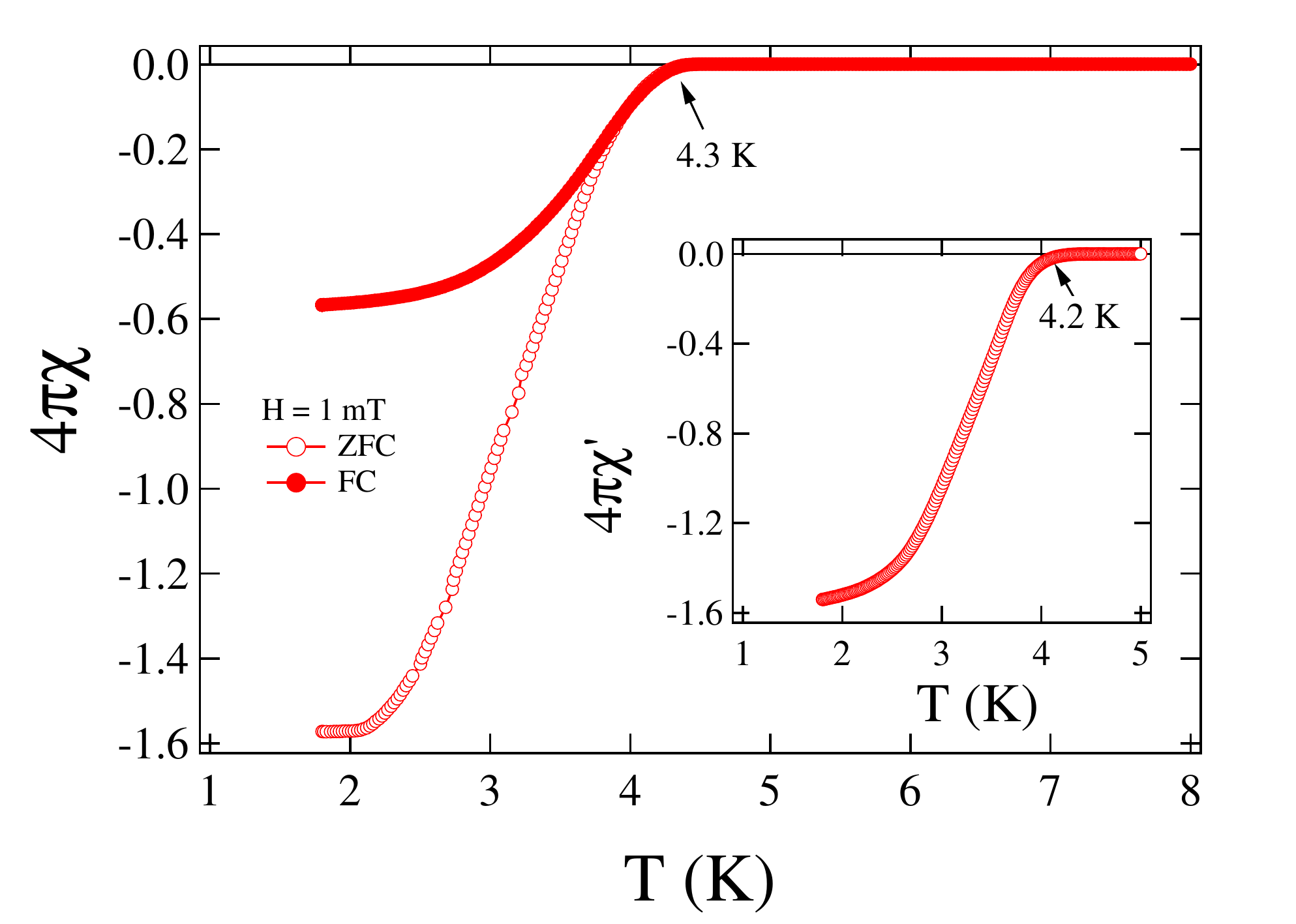}
\caption{\label{Fig6:MT}The temperature variation of dc-magnetization for Re$_{0.56}$Nb$_{0.11}$Ti$_{0.11}$Zr$_{0.11}$Hf$_{0.11}$ taken in 1 mT. Ac magnetization is shown in the inset}
\end{figure}
The superconducting state of this new material is further verified by dc and ac susceptibility measurements (Fig. \ref{Fig6:MT}), which confirm the evidence of bulk superconductivity (superconducting volume fraction is more than 100 $\%$) in the sample with a superconducting transition at 4.5 K. 
\begin{figure}
\includegraphics[width=1.0\columnwidth]{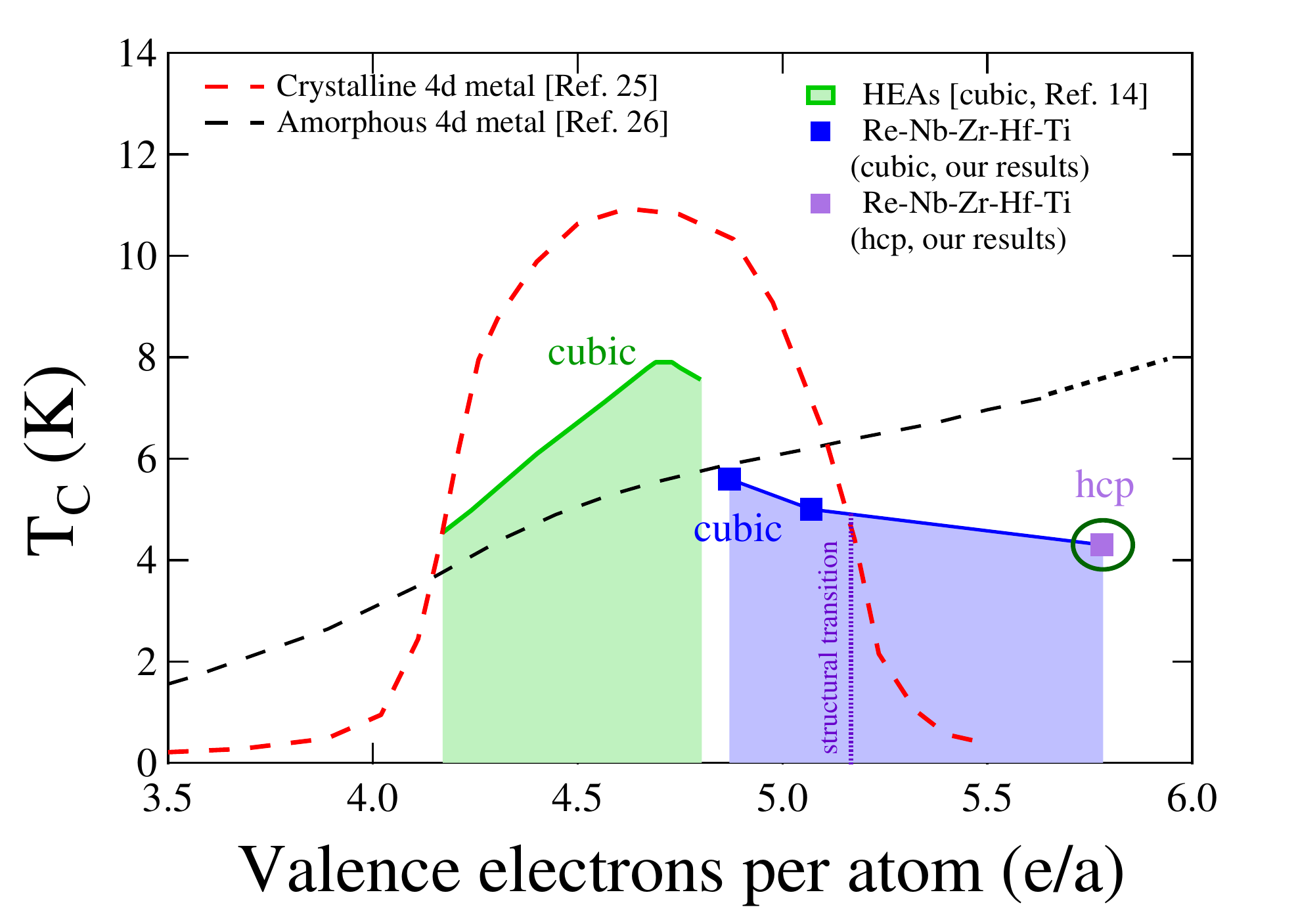}
\caption{\label{Fig7:CPLOT}Electron count (e/a) variation of the superconducting transition temperature for [Nb$_{0.67-x}$Re$_x$][TiZrHf]$_{0.33}$ materials. In comparison, crystalline 4d metals \cite{25}, amorphous 4d metals \cite{26} and the results on the effect of electron count in the Ta-Nb-Hf-Zr-Ti \cite{14} HEA system are also included here.}
\end{figure}
Fig. \ref{Fig7:CPLOT} summarizes the variation of $T_{c}$ with e/a ratio. In comparison, we have included the data of the crystalline \cite{25} and amorphous \cite{26} 4d metals. The results on the effect of electron count in the Ta-Nb-Hf-Zr-Ti \cite{14} system are also included here. Our results on electron count dependence of $T_{c}$ show a crystalline-like behaviour for the [Nb$_{0.67-x}$Re$_x$][TiZrHf]$_{0.33}$ materials. 

\begin{figure}
\includegraphics[width=1.0\columnwidth]{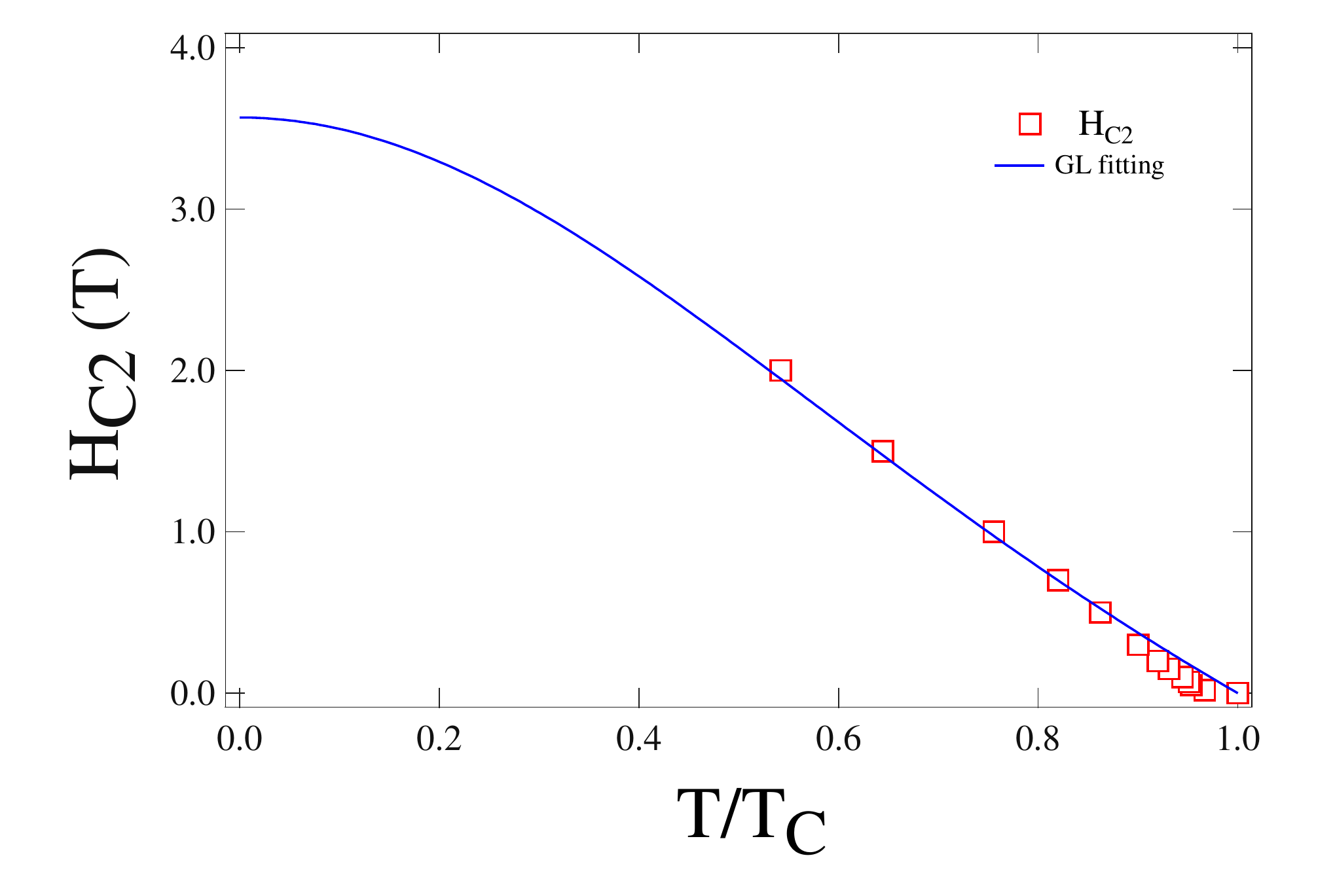}
\caption{\label{Fig8:HC2} Determination of the upper critical field ($H_{c2}$(0)) using resistivity measurements for Re$_{0.56}$Nb$_{0.11}$Ti$_{0.11}$Zr$_{0.11}$Hf$_{0.11}$. The solid line represents the Ginzburg-Landau fitting, yielding a $H_{c2}$(0) of 3.6 T.}
\end{figure}

The magnetic field variation of magnetization (M-H) highlights the type-II superconductivity in hexagonal Re$_{0.56}$Nb$_{0.11}$Ti$_{0.11}$Zr$_{0.11}$Hf$_{0.11}$. Lower critical field, $H_{c1}$(T), which is defined as the field deviating from the linear line for initial slope in the M-H curve is found to be 2.3 mT at zero temperature by the Ginzburg-Landau approximation. Detailed study on the superconducting (SC) state for the Re$_{0.56}$Nb$_{0.11}$Ti$_{0.11}$Zr$_{0.11}$Hf$_{0.11}$ material was carried out by measuring the resistivity under various applied magnetic fields. The critical temperatures are derived from the midpoint values of the superconducting transition in the resistivity measurements. The temperature variation of the upper critical field ($H_{c2}$) is shown in Fig. \ref{Fig8:HC2}. The experimental $H_{c2}$ can be described by the Ginzburg-Landau expression.
\begin{equation}
H_{c2}(T) = H_{c2}(0)\frac{(1-(T/T_{c})^{2})}{(1+(T/T_{c})^2)}
\label{eqn1:hc2}
\end{equation} 
The estimated value of the upper limit for the upper critical field, $H_{c2}$(0) is 3.6 T, and this is well below the BCS weak coupling Pauli limit ($H^{p}_{c2}$(0) = 1.84$T_{c}$, $T_{c}$ = 4.5 K gives $H^{p}_{c2}$(0) $\approx$ 8.3 T). The Ginzburg-Landau coherence length $\xi_{GL}$(0), estimated from the following relation \cite{27} is 11 nm.
\begin{equation}
H_{c2}(0) = \frac{\Phi_{0}}{2\pi\xi_{GL}^{2}}
\label{eqn2:up}
\end{equation} 
To analyze the phonon properties, electronic density of states and the nature of the superconducting state (BCS or unconventional) in this new material, we have performed the heat capacity measurement. The superconducting transition (a jump in the heat capacity data) is observed at 4.4K, confirming the bulk superconductivity. The low-temperature normal state specific heat can be well fitted using the equation $C_{P}/T$ = $\gamma$+$\beta$$T^{2}$ ($\gamma$ = Sommerfeld coefficient and $\beta$ is the lattice contribution to the specific heat) yields $\gamma$ = 2.77 mJ/mol K$^{2}$ and $\beta$ = 0.0648 mJ/mol K$^{4}$ . 
\begin{figure}
\includegraphics[width=1.0\columnwidth]{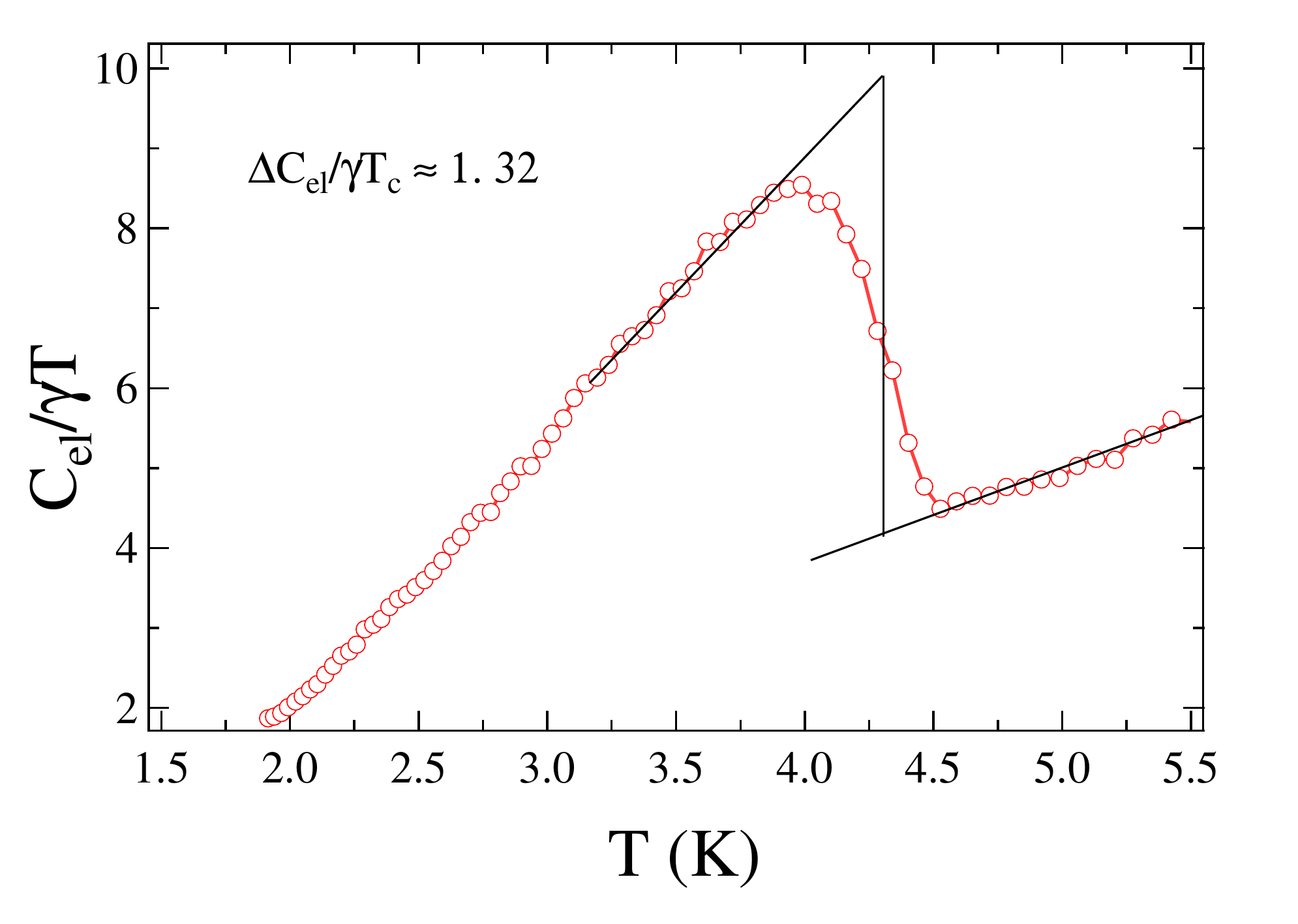}
\caption{\label{Fig9:C/T}${C_{el}}/{\gamma T}$ vs. T plot for the hexagonal Re$_{0.56}$Nb$_{0.11}$Ti$_{0.11}$Zr$_{0.11}$Hf$_{0.11}$. The ratio $\frac{\Delta C_{el}}{\gamma_{n}T_{c}}$ = 1.32 is highlighted in the figure.}
\end{figure}
The Debye temperature $\theta_{D}$ is obtained using the formula $\theta_{D} = \left(12\pi^{4}RN/5\beta\right)^{1/3}$ (4) where R is the molar gas constant (=8.314 J mol$^{-1}$ K$^{-1}$). Using N = the number of atoms per formula unit = 1, it gives $\theta_{D}$ = 310 K. The density of states values at Fermi level [$N(E_{F})$ =  $\frac{3\gamma}{\pi^{2}K_{B}^{2}}$] is calculated as 1.17 states eV$^{-1}$ f.u.$^{-1}$. The strength of the attraction between the electron and phonon, i.e., electron-phonon coupling constant can be calculated using the McMillan formula \cite{28} 

\begin{equation}
\lambda_{e-ph} = \frac{1.04+\mu^{*}ln(\theta_{D}/1.45T_{c})}{(1-0.62\mu^{*})ln(\theta_{D}/1.45T_{c})-1.04 }
\label{eqn3:ld}
\end{equation} 

By considering the Coulomb pseudopotential $\mu^{*}$ = 0.13, commonly used for many intermetallic superconductors \cite{11,15,29,30}, $\lambda_{e-ph}$ are calculated as 0.57. To know the nature of the SC state (e.g., BCS type or unconventional), the quantity of our interest is the ratio $\frac{\Delta C_{el}}{\gamma_{n}T_{c}}$. This value of this ratio is 1.43 within the BCS theory for phonon-mediated superconductivity in the weak electron-phonon coupling limit. The normalized specific heat jump is obtained as 1.32 for the present material, which is slightly lower than the BCS value in the weak coupling limit and indicates weak coupling superconductivity in the sample. 

\section{Conclusion}

In conclusion, we have synthesized a new, single phase HEA material with composition Re$_{0.56}$Nb$_{0.11}$Ti$_{0.11}$Zr$_{0.11}$Hf$_{0.11}$ by arc melting technique. Room temperature X-ray diffraction, Precession electron diffraction (PED) and precession electron diffraction tomography (PEDT) measurements are used to determine the crytal structure of this new material. Our study indicate that Re$_{0.56}$Nb$_{0.11}$Ti$_{0.11}$Zr$_{0.11}$Hf$_{0.11}$ is arranged in a hcp crystal lattice (S.G. P6$_3$/mmc) with a = b = 5.2554 (1)\text{\AA}, c = 8.5934 (1)\text{\AA}. With lowering the valence electron counts (e/a) the material shows a unique structural transition to a cubic lattice. Transport, magnetization, and thermodynamic measurements reveal that this new HEA is a type-II superconductor with the bulk superconducting transition at T$_c$=4.4 K, upper critical field $H_{c2}$ = 3.6 T. Evaluation of the low-temperature specific heat data indicates that the investigated HEA is close to a BCS-type phonon-mediated weakly coupled superconductor. Re$_{0.56}$Nb$_{0.11}$Ti$_{0.11}$Zr$_{0.11}$Hf$_{0.11}$ is the first example of a hcp superconducting high entropy alloy (HEA) and, in fact, the single ambient pressure synthesized transition metal only hcp HEA. The excellent combination of hcp structure and superconductivity in this new material, therefore offer the unique opportunity to study the exotic quantum phenomenon like superconductivity in the most desirable crystal structure in HEA family. At the same time, this will pave a new avenue for the exploration of future generation superconducting materials.

\section{Acknowledgments}

R.~P.~S.\ acknowledges Science and Engineering Research Board, Government of India for the Ramanujan Fellowship through Grant No. SR/S2/RJN-83/2012. S.~M.\ acknowledges Science and Engineering Research Board, Government of India for the NPDF Fellowship (PDF/2016/000348).

\end{document}